\documentclass[12pt]{article}%
\usepackage{amsmath}
\usepackage{amsfonts}
\usepackage{amssymb}
\usepackage{graphicx}%
\begin{document}
\begin{titlepage}
\begin{flushright}
CAMS/04-07\\
\end{flushright}
\vspace{.3cm}
\begin{center}
\renewcommand{\thefootnote}{\fnsymbol{footnote}}
{\Large\bf Matrix Gravity and Massive Colored Gravitons} \vskip20mm
{\large\bf{Ali H.
Chamseddine \footnote{email: chams@aub.edu.lb}}}\\
\renewcommand{\thefootnote}{\arabic{footnote}}
\vskip2cm {\it Center for Advanced Mathematical Sciences (CAMS)
and\\
Physics Department, American University of Beirut, Lebanon.\\}
\end{center}
\vfill
\begin{center}
{\bf Abstract}
\end{center}
\vskip 1cm We formulate a theory of gravity with a matrix-valued
complex vierbein based on the $SL(2N,\mathbb{C})\otimes
SL(2N,\mathbb{C})$ gauge symmetry. The theory is metric
independent, and before symmetry breaking all fields are massless.
The symmetry is broken spontaneously and all gravitons
corresponding to the broken generators acquire masses. If the
symmetry is broken to $SL(2,\mathbb{C})$ then the spectrum would
correspond to one massless graviton coupled to $2N^2 -1$ massive
gravitons. A novel feature is the way the fields corresponding to
non-compact generators acquire kinetic energies with correct
signs. Equally surprising is the way Yang-Mills gauge fields
acquire their correct kinetic energies through the coupling to the
non-dynamical antisymmetric components of the vierbeins.
\end{titlepage}

\section{Introduction}

The basic interactions of the gravitational field could be easily deduced by
promoting the global Lorentz invariance of the Dirac equation to a local one.
This simple, yet powerful observation, was first made by Weyl \cite{Weyl} and
Cartan \cite{Cartan}, and later developed by Utiyama \cite{Utiyama} and Kibble
\cite{Kibble}, to formulate Einstein's theory of gravity as a gauge theory of
$SL(2,\mathbb{C}).$ The crucial property that guarantees masslessness of the
graviton is invariance under diffeomorphisms. Dynamical theories of gauge
fields are based on internal symmetries. The space-time symmetry group
$SL(2,\mathbb{C})$ and internal symmetry groups are distinct and not unified.
According to the Coleman-Mandula theorem \cite{Coleman} the only possible
extensions to space-time symmetries are internal ones. There is, however, the
exception of supersymmetries which are based on extending the Poincare
symmetry by grading the algebra.

Some time ago, Isham, Salam and Strathdee \cite{ISS} proposed a marriage of
$SL(2,\mathbb{C})$ and $SU(3)$ symmetries to describe the interaction of
massive spin $2^{+}$ nonets by gauging the unified $SL(6,\mathbb{C})$ group.
Their construction is analogous to the first order formalism of gravity where
a matrix-valued connection vector is introduced\ as a conjugate variable to
the matrix-valued gauge field. This method insures that degrees of freedom
associated with the non-compact components of $SL(6,\mathbb{C})$ do not
propagate. This theory was intended to describe strong interactions and the
space-time metric was taken to be flat Minkowski.

To unify gravity with internal symmetries, the mechanism introduced in
\cite{ISS} can serve as a starting point, with the graviton promoted to become
matrix-valued. To insure diffeomorphism invariance of matrix-valued
expressions, no metric is introduced from outside and the action is required
to be a four-form on a four-dimensional manifold. The natural group to
consider that marries the $SL(2,\mathbb{C})$ symmetry and the internal $U(N)$
symmetry is $SL(2N,\mathbb{C}).$ The $8N^{2}-2$ generators of
$SL(2N,\mathbb{C})$ are formed from the tensor products of the generators of
$SL(2,\mathbb{C})$ and generators of $U(N)$. An attempt to formulate matrix
gravity based on the metric approach tensored to $U(N)$ matrices was
considered by Avramidi \cite{Avramidi}. However, in \cite{Avramidi} all
gravitons have unbroken $U(N)$ symmetry and are all massless which is
non-physical. More recently he obtained a unique gravity action based on the
spectral expansion of non-Laplace type operator \cite{avramidi2}. That
formulation differs from the approach considered here as we take the mixing
between the $SL(2,\mathbb{C})$ space-time symmetry and the $U(N)$ internal
symmetry to be non-trivial, and where after symmetry breaking, all gravitons
except one become massive. As a starting point, the vierbein can be taken as a
one form transforming under the vector representation of $SL(2N,\mathbb{C})$
symmetry. In the Dirac basis the vierbein will have two components, a vector
and pseudo-vector. When expressed in terms of components, this corresponds to
a complex matrix-valued vierbein. It turns out that finding a Lagrangian with
physically acceptable propagators for all components of the complex
matrix-valued vierbein, is not possible within this setting. The reason is
that the vector and pseudo-vector components of the vierbein will get kinetic
energies of opposite signs. In analogy with what was done in reference
\cite{ISS}, the correct strategy to adopt is to extend the gauge group to
$SL(2N,\mathbb{C})\otimes SL(2N,\mathbb{C})$ . This will remedy the problem of
getting the correct signs for the kinetic energies. It will allow us to impose
torsion constraints which could then be used to solve for the spin-connection
parts of the gauge fields in terms of the vierbeins. It will also be necessary
to introduce scalar fields that spontaneously break the symmetry
$SL(2N,\mathbb{C})\otimes SL(2N,\mathbb{C})$ to $SL(2,\mathbb{C})$ \cite{ISS}.
This will then permit the coupling of the gravitational sector to matter.
Diffeomorphism invariance can only protect one metric, a scalar under the
$SU(N)$ factors, from getting a mass. The massive gravitons, except for one
singlet, are in a representation of $SU(N)\times SU(N)$ and are therefore
colored. Here we shall not worry about the consistency of massive spin-2
theories \cite{Vandam}, \cite{Deser}, \cite{Damour}, as this problem has been
recently resolved as shown in \cite{Georgi} and \cite{spontaneous}.The plan of
this paper is as follows. In section 2 the gauge fields with their
transformations and curvatures are introduced, and torsion free conditions are
imposed. In section 3 a metric independent gauge invariant Lagrangian is
constructed, and Higgs fields necessary for breaking the symmetry
spontaneously are used. In section 4 the spectrum of the Lagrangian is
analyzed by deriving the quadratic terms perturbatively. Section 5 is the conclusion.

\section{Matrix Gravity}

As a starting point we introduce the group $SL(2N,\mathbb{C})\otimes
SL(2N,\mathbb{C})$ defined as the complex extension of $SL(2N,\mathbb{C})$
with complex gauge parameters $\Omega.$ This has the following representation
in the Dirac basis\footnote{We use the same notation and set of Dirac gamma
matrices as reference \cite{rasheed} except for $\gamma_{5}$ which is replaced
here with $-i\gamma_{5}.$}:%
\begin{align*}
\Omega &  =\exp\omega,\quad\\
\omega &  =P_{+}\left(  \omega_{l}+\frac{i}{4}l^{ab}\sigma_{ab}\right)
+P_{-}\left(  \omega_{r}+\frac{i}{4}r^{ab}\sigma_{ab}\right)  ,
\end{align*}
where, $\det\Omega=1,$ and
\begin{align*}
\omega_{l}  &  =\omega_{1}+i\omega_{2},\\
\qquad\omega_{r}  &  =\omega_{1}^{^{\prime}}+i\omega_{2}^{^{\prime}},\qquad\\
P_{\pm}  &  =\frac{1}{2}\left(  1\pm\gamma_{5}\right)  .
\end{align*}
The gauge parameters $\omega_{1},$ $\omega_{2},$ $\omega_{1}^{^{\prime}},$
$\omega_{2}^{^{\prime}},$ $l^{ab},$ and $r^{ab}$ are Hermitian $N\times N$
\ matrices. The gauge fields are one forms
\[
A=A_{\mu}dx^{\mu},\quad
\]
where $A_{\mu}$ are defined by%
\[
A_{\mu}=P_{+}\left(  a_{\mu}+\frac{i}{4}B_{\mu}^{ab}\sigma_{ab}\right)
+P_{-}\left(  b_{\mu}+\frac{i}{4}C_{\mu}^{ab}\sigma_{ab}\right)  ,
\]
and satisfy the conditions $Tr\left(  P_{\pm}A_{\mu}\right)  =0.$ The
$SU(N)\times SU(N)$ gauge fields $a_{\mu}$ and $b_{\mu}$ are complex
\begin{align*}
a_{\mu}  &  =a_{1\mu}+ia_{2\mu},\\
\qquad b_{\mu}  &  =b_{1\mu}+ib_{2\mu},
\end{align*}
and the component gauge fields $a_{1\mu},$ $a_{2\mu},$ $b_{1\mu},$ $b_{2\mu},$
$B_{\mu}^{ab}$ and $C_{\mu}^{ab}$ are taken to be Hermitian $N\times N$
matrices subject to the conditions%
\[
Tr\left(  a_{\mu}\right)  =0=Tr\left(  b_{\mu}\right)  .
\]
Under a gauge transformation, the gauge fields transform as
\[
A_{\mu}\rightarrow\Omega A_{\mu}\Omega^{-1}+\Omega\partial_{\mu}\Omega^{-1}.
\]
The complex conjugate of the group element $\Omega$ is defined by
\[
\overline{\Omega}=\gamma_{0}\Omega^{\dagger}\gamma_{0}=\exp\overline{\omega},
\]
so that
\[
\overline{\omega}=P_{+}\left(  \overline{\omega_{r}}-\frac{i}{4}r^{ab}%
\sigma_{ab}\right)  +P_{-}\left(  \overline{\omega_{l}}-\frac{i}{4}%
l^{ab}\sigma_{ab}\right)  ,
\]
where
\[
\overline{\omega_{l}}=\omega_{1}-i\omega_{2},\qquad\overline{\omega_{r}%
}=\omega_{1}^{^{\prime}}-i\omega_{2}^{^{\prime}}.
\]
The complex conjugate gauge fields are then%
\[
\overline{A}_{\mu}=P_{+}\left(  \overline{b}_{\mu}-\frac{i}{4}C_{\mu}%
^{ab}\sigma_{ab}\right)  +P_{-}\left(  \overline{a}_{\mu}-\frac{i}{4}B_{\mu
}^{ab}\sigma_{ab}\right)  ,
\]
where
\[
\overline{a}_{\mu}=a_{1\mu}-ia_{2\mu},\qquad\overline{b}_{\mu}=b_{1\mu
}-ib_{2\mu},
\]
transform as
\[
\overline{A}_{\mu}\rightarrow\overline{\Omega}^{-1}\overline{A}_{\mu}%
\overline{\Omega}-\overline{\Omega}^{-1}\partial_{\mu}\overline{\Omega}.
\]
Next, we introduce the two matrix valued vierbeins one forms
\[
L=L_{\mu}dx^{\mu},\qquad L^{^{\prime}}=L_{\mu}^{^{\prime}}dx^{\mu},
\]
where
\begin{align*}
L_{\mu}  &  =\left(  P_{+}E_{\mu}^{a}+P_{-}F_{\mu}^{a}\right)  \gamma_{a},\\
L_{\mu}^{^{\prime}}  &  =\left(  P_{+}F_{\mu}^{^{\prime}a}+P_{-}E_{\mu
}^{^{\prime}a}\right)  \gamma_{a},
\end{align*}
and these transform under product representations of the two groups
$SL(2N,\mathbb{C}):$%
\begin{align*}
L_{\mu}  &  \rightarrow\Omega L_{\mu}\overline{\Omega}\\
L_{\mu}^{^{\prime}}  &  \rightarrow\overline{\Omega}^{-1}L_{\mu}^{^{\prime}%
}\Omega^{-1}%
\end{align*}
Notice that
\[
\overline{L}_{\mu}=\gamma_{0}L_{\mu}^{\dagger}\gamma_{0}=L_{\mu}%
\]
and
\[
\overline{L_{\mu}^{^{\prime}}}=\gamma_{0}L_{\mu}^{^{\prime}\dagger}\gamma
_{0}=L_{\mu}^{^{\prime}}%
\]
and the combinations $L_{\mu}L_{\nu}^{^{\prime}}$ and $L_{\mu}^{^{\prime}%
}L_{\nu}$ transform as%
\begin{align*}
L_{\mu}L_{\nu}^{^{\prime}}  &  \rightarrow\Omega L_{\mu}L_{\nu}^{^{\prime}%
}\Omega^{-1},\\
L_{\mu}^{^{\prime}}L_{\nu}  &  \rightarrow\overline{\Omega}^{-1}L_{\mu
}^{^{\prime}}L_{\nu}\overline{\Omega},
\end{align*}
It is instructive to write down the transformations for the component fields
$E_{\mu}^{a},$ $E_{\mu}^{^{\prime}a},$ $F_{\mu}^{a}$ and $F_{\mu}^{^{\prime}%
a}$ ,%
\begin{align*}
\delta E_{\mu}^{a}  &  =\left\{  \alpha_{1},E_{\mu}^{a}\right\}  +i\left[
\alpha_{2},E_{\mu}^{a}\right]  -\frac{1}{2}\left\{  l^{ab},E_{\mu b}\right\}
-\frac{i}{2}\left[  \widetilde{l}^{ab},E_{\mu b}\right]  ,\\
\delta F_{\mu}^{a}  &  =\left\{  \beta_{1},F_{\mu}^{a}\right\}  +i\left[
\beta_{2},F_{\mu}^{a}\right]  -\frac{1}{2}\left\{  r^{ab},F_{\mu b}\right\}
+\frac{i}{2}\left[  \widetilde{r}^{ab},F_{\mu b}\right]  ,\\
\delta E_{\mu}^{^{\prime}a}  &  =\left\{  \alpha_{1},E_{\mu}^{^{\prime}%
a}\right\}  +i\left[  \alpha_{2},E_{\mu}^{^{\prime}a}\right]  -\frac{1}%
{2}\left\{  l^{ab},E_{\mu b}^{^{\prime}}\right\}  +\frac{i}{2}\left[
\widetilde{l}^{ab},E_{\mu b}^{^{\prime}}\right]  ,\\
\delta F_{\mu}^{^{\prime}a}  &  =\left\{  \beta_{1},F_{\mu}^{^{\prime}%
a}\right\}  +i\left[  \beta_{2},F_{\mu}^{^{\prime}a}\right]  -\frac{1}%
{2}\left\{  r^{ab},F_{\mu b}^{^{\prime}}\right\}  -\frac{i}{2}\left[
\widetilde{r}^{ab},F_{\mu b}^{^{\prime}}\right]  ,
\end{align*}
where
\[
\widetilde{l}^{ab}=\frac{1}{2}\epsilon^{abcd}l_{cd},\qquad\widetilde{r}%
^{ab}=\frac{1}{2}\epsilon^{abcd}r_{cd}.
\]
It is clear that one can only identify the fields $E_{\mu}^{^{\prime}a}%
=E_{\mu}^{a}$ and $F_{\mu}^{^{\prime}a}=F_{\mu}^{a}$ in the special case of
$SL(2,\mathbb{C})\otimes SL(2,\mathbb{C})$ \cite{AC} as then, the last terms
in the previous transformations, which occur with opposite signs vanish. It
was shown in reference \cite{AC} that, in this $N=1$ case, it is not essential
to consider $SL(2,\mathbb{C})\otimes SL(2,\mathbb{C})$ and it is sufficient to
have the group $SL(2,\mathbb{C})$ only. It is instructive, however, to work
out the details in this case to illustrate the generation of mass through the
Higgs mechanism.

To avoid doubling the dynamical degrees of freedom, we shall impose
constraints on the system so that only one pair of $E_{\mu}^{a}$ and $F_{\mu
}^{a}$ propagate.

The curvature of the one-forms $A$ and $\overline{A}$ are defined by
\begin{align*}
F  &  =dA+A^{2}=\frac{1}{2}F_{\mu\nu}dx^{\mu}\wedge dx^{\nu},\\
\overline{F}  &  =d\overline{A}-\overline{A}^{2}=\frac{1}{2}\overline{F}%
_{\mu\nu}dx^{\mu}\wedge dx^{\nu},
\end{align*}
which give the following field strengths
\begin{align*}
F_{\mu\nu}  &  =\partial_{\mu}A_{\nu}-\partial_{\nu}A_{\mu}+\left[  A_{\mu
},A_{\nu}\right]  ,\\
\overline{F}_{\mu\nu}  &  =\partial_{\mu}\overline{A}_{\nu}-\partial_{\nu
}\overline{A}_{\mu}-\left[  \overline{A}_{\mu},\overline{A}_{\nu}\right]  .
\end{align*}
These transform as
\begin{align*}
F_{\mu\nu}  &  \rightarrow\Omega F_{\mu\nu}\Omega^{-1},\\
\overline{F}_{\mu\nu}  &  \rightarrow\overline{\Omega}^{-1}\overline{F}%
_{\mu\nu}\overline{\Omega}.
\end{align*}
Next we introduce the analogue of torsion forms%
\begin{align*}
T  &  =dL+AL+LA^{^{\prime}}=\frac{1}{2}T_{\mu\nu}dx^{\mu}\wedge dx^{\nu},\\
T^{^{\prime}}  &  =dL^{^{\prime}}+AL^{^{\prime}}+L^{^{\prime}}A^{^{\prime}%
}=\frac{1}{2}T_{\mu\nu}^{^{\prime}}dx^{\mu}\wedge dx^{\nu}.
\end{align*}
The tensors $T_{\mu\nu}$ and $T_{\mu\nu}^{^{\prime}}$ can be expanded in the
Dirac basis%
\begin{align*}
T_{\mu\nu}  &  =\left(  P_{+}T_{\mu\nu}^{a}\left(  E\right)  +P_{-}T_{\mu\nu
}^{a}\left(  F\right)  \right)  \gamma_{a},\\
T_{\mu\nu}^{^{\prime}}  &  =\left(  P_{+}T_{\mu\nu}^{a}\left(  F^{^{\prime}%
}\right)  +P_{-}T_{\mu\nu}^{a}\left(  E^{^{\prime}}\right)  \right)
\gamma_{a},
\end{align*}
where
\begin{align*}
T_{\mu\nu}^{a}\left(  E\right)   &  =\nabla_{\mu}E_{\nu}^{a}-\nabla_{\nu
}E_{\mu}^{a},\\
T_{\mu\nu}^{a}\left(  F\right)   &  =\nabla_{\mu}F_{\nu}^{a}-\nabla_{\nu
}F_{\mu}^{a},
\end{align*}
and similarly for $T_{\mu\nu}^{a}\left(  E^{^{\prime}}\right)  $ and
$T_{\mu\nu}^{a}\left(  F^{^{\prime}}\right)  .$ The covariant derivatives are
given by
\begin{align*}
\nabla_{\mu}E_{\nu}^{a}  &  =\partial_{\mu}E_{\nu}^{a}+\left\{  a_{\mu}%
^{1},E_{\nu}^{a}\right\}  +i\left[  a_{\mu}^{2},E_{\nu}^{a}\right]  +\frac
{1}{2}\left\{  B_{\mu}^{ab},E_{\nu b}\right\}  +\frac{i}{2}\left[
\widetilde{B}_{\mu}^{ab},E_{\nu b}\right]  ,\\
\nabla_{\mu}E_{\nu}^{^{\prime}a}  &  =\partial_{\mu}E_{\nu}^{^{\prime}%
a}+\left\{  a_{\mu}^{1},E_{\nu}^{^{\prime}a}\right\}  +i\left[  a_{\mu}%
^{2},E_{\nu}^{^{\prime}a}\right]  +\frac{1}{2}\left\{  B_{\mu}^{ab},E_{\nu
b}^{^{\prime}}\right\}  -\frac{i}{2}\left[  \widetilde{B}_{\mu}^{ab},E_{\nu
b}^{^{\prime}}\right]  ,\\
\nabla_{\mu}F_{\nu}^{a}  &  =\partial_{\mu}F_{\nu}^{a}+\left\{  b_{\mu}%
^{1},F_{\nu}^{a}\right\}  +i\left[  b_{\mu}^{2},F_{\nu}^{a}\right]  +\frac
{1}{2}\left\{  C_{\mu}^{ab},F_{\nu b}\right\}  -\frac{i}{2}\left[
\widetilde{C}_{\mu}^{ab},F_{\nu b}\right]  ,\\
\nabla_{\mu}F_{\nu}^{^{\prime}a}  &  =\partial_{\mu}F_{\nu}^{^{\prime}%
a}+\left\{  b_{\mu}^{1},F_{\nu}^{^{\prime}a}\right\}  +i\left[  b_{\mu}%
^{2},F_{\nu}^{^{\prime}a}\right]  +\frac{1}{2}\left\{  C_{\mu}^{ab},F_{\nu
b}^{^{\prime}}\right\}  +\frac{i}{2}\left[  \widetilde{C}_{\mu}^{ab},F_{\nu
b}^{^{\prime}}\right]  ,
\end{align*}
where we have defined
\begin{align*}
\widetilde{B}_{\mu}^{ab}  &  =\frac{1}{2}\epsilon^{abcd}B_{\mu cd},\\
\widetilde{C}_{\mu}^{ab}  &  =\frac{1}{2}\epsilon^{abcd}C_{\mu cd}.
\end{align*}
From the transformation properties of $E_{\mu}^{a}$, $F_{\mu}^{a},$ $E_{\mu
}^{^{\prime}a}$ and $F_{\mu}^{^{\prime}a}$ it is clear that it is not possible
to make the identification
\begin{align*}
E_{\mu}^{^{\prime}a}  &  =E_{\mu}^{a},\qquad\\
F_{\mu}^{^{\prime}a}  &  =F_{\mu}^{a},
\end{align*}
without breaking the $SL(2N,\mathbb{C})\otimes SL(2N,\mathbb{C})$ to
$SL(2,\mathbb{C}).$ In analogy with the Cartan formulation of gravity, we
shall impose the following zero torsion constraints%
\begin{align*}
T_{\mu\nu}  &  =0,\\
T_{\mu\nu}^{^{\prime}}  &  =0.
\end{align*}
The number of independent components in $T_{\mu\nu}^{a}\left(  E\right)  ,$
$T_{\mu\nu}^{a}\left(  F\right)  $ or $T_{\mu\nu}^{a}\left(  E^{^{\prime}%
}\right)  $ and $T_{\mu\nu}^{a}\left(  F^{^{\prime}}\right)  ,$ match the
number of independent components in $B_{\mu}^{ab}$ and $C_{\mu}^{ab}.$ We can
then use the $T_{\mu\nu}=0$ constraints to solve for $B_{\mu}^{ab}$ in terms
of the fields $a_{\mu},$ $E_{\mu}^{a}$ and for $C_{\mu}^{ab}$ in terms of
$b_{\mu}$, $F_{\mu}^{a}.$ Alternatively we can use the constraint $T_{\mu\nu
}^{^{\prime}}=0$ to solve for $B_{\mu}^{ab}$ in terms of the fields $a_{\mu}$,
$E_{\mu}^{^{\prime}a}$ and for $C_{\mu}^{ab}$ in terms of $b_{\mu},$ $F_{\mu
}^{^{\prime}a}$. Therefore, equating the two sets of solutions give a
complicated relation between $a_{\mu}$, $E_{\mu}^{a}$, $E_{\mu}^{^{\prime}a},$
as well as another relation between $b_{\mu}$, $F_{\mu}^{a}$ and $F_{\mu
}^{^{\prime}a}.$ These relations being matrix equations could only be solved
perturbatively. The important point is that the constraints reduce the number
of independent fields from two sets of complex matrix-valued vierbeins to one
set. After writing the action, we shall analyze the dynamical degrees of freedom.

\bigskip The components of the curvature $F_{\mu\nu}$ are given by%
\[
F_{\mu\nu}=P_{+}\left(  a_{\mu\nu}+\frac{i}{4}B_{\mu\nu}^{ab}\sigma
_{ab}\right)  +P_{-}\left(  b_{\mu\nu}+\frac{i}{4}C_{\mu\nu}^{ab}\sigma
_{ab}\right)  ,
\]
where
\begin{align*}
a_{\mu\nu}  &  =\partial_{\mu}a_{\nu}-\partial_{\nu}a_{\mu}-\frac{1}{8}\left[
B_{\mu}^{ab},B_{\nu ab}\right]  -\frac{1}{16}\epsilon_{abcd}\left[  B_{\mu
}^{ab},B_{\nu}^{cd}\right]  ,\\
b_{\mu\nu}  &  =\partial_{\mu}b_{\nu}-\partial_{\nu}b_{\mu}-\frac{1}{8}\left[
C_{\mu}^{ab},C_{\nu ab}\right]  -\frac{1}{16}\epsilon_{abcd}\left[  C_{\mu
}^{ab},C_{\nu}^{cd}\right]  ,\\
B_{\mu\nu}^{\;ab}  &  =\partial_{\mu}B_{\nu}^{\;ab}+\frac{1}{2}\left\{
B_{\mu}^{\;ac},B_{\nu c}^{\;\,\;b}\right\}  +i\left[  B_{\mu}^{\;ab},a_{\nu
}^{2}\right]  +i\left[  \widetilde{B}_{\mu}^{\;ab},a_{\nu}^{1}\right]
-\mu\leftrightarrow\nu,\\
C_{\mu\nu}^{\;ab}  &  =\partial_{\mu}C_{\nu}^{\;ab}-\partial_{\nu}C_{\mu
}^{\;ab}+\frac{1}{2}\left\{  C_{\mu}^{\;ac},C_{\nu c}^{\;\,\;b}\right\}
+i\left[  C_{\mu}^{\;ab},b_{\nu}^{2}\right]  +i\left[  \widetilde{C}_{\mu
}^{\;ab},b_{\nu}^{1}\right]  -\mu\leftrightarrow\nu.
\end{align*}
The complex conjugate gauge field strength $\overline{F}$ are not independent
and are given by%
\[
\overline{F}_{\mu\nu}=P_{+}\left(  \overline{b}_{\mu\nu}-\frac{i}{4}C_{\mu\nu
}^{ab}\sigma_{ab}\right)  +P_{-}\left(  \overline{a}_{\mu\nu}-\frac{i}%
{4}B_{\mu\nu}^{ab}\sigma_{ab}\right)  .
\]

\section{\bigskip Metric Independent Lagrangian}

The model we want to construct should also include the Einstein gravitational
field. Therefore, we will not use the metric given on the manifold $M$, but
make the requirement that it should arise dynamically from the vierbeins. This
will impose the very strong constraint that every term in the Lagrangian
should be a four-form, to insure diffeomorphism invariance. With the fields
introduced so far, it is possible to build a limited number of gauge invariant
terms which are also four-forms. These are
\[
\frac{1}{4}%
{\displaystyle\int\limits_{M}}
Tr\left(  i\left(  \alpha+\beta\gamma_{5}\right)  LL^{^{\prime}}F+i\left(
\overline{\alpha}+\overline{\beta}\gamma_{5}\right)  L^{^{\prime}}%
L\overline{F}+\left(  i\lambda+\gamma_{5}\eta\right)  LL^{^{\prime}%
}LL^{^{\prime}}\right)  ,
\]
where $\alpha=\alpha_{1}+i\alpha_{2},$ $\overline{\alpha}=\alpha_{1}%
-i\alpha_{2},$ $\beta=\beta_{1}+i\beta_{2},$ $\overline{\beta}=\beta
_{1}-i\beta_{2}$ are coupling constants.

Before proceeding, it is instructive to consider the simple case of $N=1$.
Here the $SL(2,\mathbb{C})$ conditions imply that $a_{\mu}=0=b_{\mu},$ and the
two conditions $T_{\mu\nu}=$ $T_{\mu\nu}^{^{\prime}}=0$ could be solved to
give
\begin{align*}
E_{\mu}^{^{\prime}a}  &  =E_{\mu}^{a}=e_{\mu}^{a},\\
\qquad F_{\mu}^{^{\prime}a}  &  =F_{\mu}^{a}=f_{\mu}^{a}\\
B_{\mu}^{ab}  &  =\frac{1}{2}e^{\nu a}e^{\rho b}\left(  \Omega_{\mu\nu\rho
}\left(  e\right)  -\Omega_{\nu\rho\mu}\left(  e\right)  +\Omega_{\rho\mu\nu
}\left(  e\right)  \right)  \equiv\omega_{\mu}^{ab}\left(  e\right)  ,\\
C_{\mu}^{ab}  &  =\frac{1}{2}f^{\nu a}f^{\rho b}\left(  \Omega_{\mu\nu\rho
}\left(  f\right)  -\Omega_{\nu\rho\mu}\left(  f\right)  +\Omega_{\rho\mu\nu
}\left(  f\right)  \right)  \equiv\omega_{\mu}^{ab}\left(  f\right)  ,\\
\Omega_{\mu\nu\rho}\left(  e\right)   &  =\left(  \partial_{\mu}e_{\nu}%
^{c}-\partial_{\nu}e_{\mu}^{c}\right)  e_{\rho c},\\
\qquad\Omega_{\mu\nu\rho}\left(  f\right)   &  =\left(  \partial_{\mu}f_{\nu
}^{c}-\partial_{\nu}f_{\mu}^{c}\right)  f_{\rho c}.
\end{align*}
In this special case it is possible to identify the fields appearing in $L$
and $L^{^{\prime}}$ as was shown in \cite{AC}, because it becomes possible to
define an operation of charge conjugation that transforms $L$ to $L^{^{\prime
}}.$ With these simplifications the action reduce to%
\begin{align*}
&  -\frac{1}{2}%
{\displaystyle\int\limits_{M}}
d^{4}x\epsilon^{\mu\nu\kappa\lambda}\left(  \left(  \left(  \alpha_{2}%
-\beta_{1}\right)  e_{\mu a}e_{\nu b}+\frac{1}{2}\left(  \alpha_{1}+\beta
_{2}\right)  \epsilon_{abcd}e_{\mu}^{c}e_{\nu}^{d}\right)  B_{\kappa\lambda
}^{ab}\right. \\
&  \qquad\qquad+\left(  \left(  \alpha_{2}+\beta_{1}\right)  f_{\mu a}f_{\nu
b}-\frac{1}{2}\left(  \alpha_{1}-\beta_{2}\right)  \epsilon_{abcd}f_{\mu}%
^{c}f_{\nu}^{d}\right)  C_{\kappa\lambda}^{ab}\\
&  \qquad\qquad\left.  +\epsilon_{abcd}\left(  \left(  \lambda-\eta\right)
e_{\mu}^{a}e_{\nu}^{b}e_{\kappa}^{c}e_{\lambda}^{d}+\left(  \lambda
+\eta\right)  f_{\mu}^{a}f_{\nu}^{b}f_{\kappa}^{c}f_{\lambda}^{d}\right)
\right)  .
\end{align*}
It is clear that this action describes two non-interacting vierbeins, and what
is needed is to add to the action mixing terms to give mass to one of the
metrics. This, however, is not possible without breaking the symmetry
spontaneously from $SL(2,\mathbb{C})\otimes SL(2,\mathbb{C})$ to
$SL(2,\mathbb{C})$. \ Such program was carried many years ago in \cite{CSS}
for strong supergravity based on spontaneously breaking the graded
orthosymplectic gauge symmetry $OSP(2,2;1)\times OSP(2,2;1)$ to
$SL(2,\mathbb{C}).$ We shall return to this simple example of $N=1$ when we
analyze the spectrum.

Introduce the Higgs fields $H$ and $H^{^{\prime}}$ which transform,
respectively, as $L$ and $L^{^{\prime}}:$
\begin{align*}
H  &  \rightarrow\Omega H\overline{\Omega},\\
H^{^{\prime}}  &  \rightarrow\overline{\Omega}^{-1}H^{^{\prime}}\Omega^{-1},
\end{align*}
and subject to the reality conditions $H=\overline{H},$ $H^{^{\prime}%
}=\overline{H^{^{\prime}}}.$ The component forms of $H$ and $H^{^{\prime}}$are
given by
\begin{align*}
H  &  =\left(  h_{1}+\gamma_{5}h_{2}+\frac{1}{4}h^{ab}\sigma_{ab}\right)  ,\\
H^{^{\prime}}  &  =\left(  h_{1}^{^{\prime}}+\gamma_{5}h_{2}^{^{\prime}}%
+\frac{1}{4}h^{^{\prime}ab}\sigma_{ab}\right)  ,
\end{align*}
where all the components are real. However, in performing calculations it
proves easier to decompose $H$ and $H^{^{\prime}}$ in terms of their chiral
components%
\begin{align*}
H  &  =P_{+}\left(  h+\frac{1}{4}h_{+}^{ab}\sigma_{ab}\right)  +P_{+}\left(
h^{\ast}+\frac{1}{4}h_{-}^{ab}\sigma_{ab}\right)  ,\\
H  &  =P_{+}\left(  h^{^{\prime}}+\frac{1}{4}h_{+}^{^{\prime}ab}\sigma
_{ab}\right)  +P_{+}\left(  h^{^{\prime}\ast}+\frac{1}{4}h_{-}^{^{\prime}%
ab}\sigma_{ab}\right)  ,
\end{align*}
where $h_{ab}=h_{+ab}+h_{-ab}$ the sum of the self-dual and anti self-dual
parts:
\begin{align*}
h_{\pm ab}  &  =\pm\frac{i}{2}\epsilon_{abcd}h_{\pm}^{cd},\qquad
h=h_{1}+ih_{2},\\
h_{\pm ab}^{^{\prime}}  &  =\pm\frac{i}{2}\epsilon_{abcd}h_{\pm}^{^{\prime}%
cd},\qquad h=h_{1}+ih_{2}.
\end{align*}
The fact that the scalar field $H$ transforms like $L$ and $H^{^{\prime}}$
transforms like $L^{^{\prime}}$ allows us to form new combinations. The choice
of terms is limited because of the constraint that every term should be a
four-form. The simplest allowed combinations are
\[
\frac{1}{4}%
{\displaystyle\int\limits_{M}}
Tr\left(  \left(  i\tau+\gamma_{5}\xi\right)  LH^{^{\prime}}LH^{^{\prime}%
}LL^{^{\prime}}+\left(  i\rho+\gamma_{5}\chi\right)  HL^{^{\prime}%
}HL^{^{\prime}}LL^{^{\prime}}\right)  .
\]
To the above, we can add kinetic terms to the field $H$ in the form
\[%
{\displaystyle\int\limits_{M}}
Tr\left(  \left(  a+b\gamma_{5}\right)  \left(  \nabla H\nabla H^{^{\prime}%
}LL^{^{\prime}}+\nabla H^{^{\prime}}\nabla HL^{^{\prime}}L\right)  \right)  ,
\]
where
\begin{align*}
\nabla H  &  =dH+AH+H\overline{A},\\
\nabla H^{^{\prime}}  &  =dH^{^{\prime}}-AH^{\prime}-H^{^{\prime}}%
\overline{A.}%
\end{align*}
Alternatively, it is possible to use the method of non-linear realization
\cite{CWZ} to constrain the Higgs fields by gauge invariant conditions of the
form%
\begin{align*}
Tr\left(  P_{\pm}\left(  HH^{^{\prime}}\right)  ^{n}\right)   &  =c_{\pm
n},\qquad n=1,2,\cdots,\\
Tr\left(  P_{\pm}H\nabla_{\mu}H^{^{\prime}}\right)   &  =0,
\end{align*}
where $c_{n}$ are constants. In this way there will be no need to add a
potential for the Higgs fields, but instead solve the constraints to eliminate
all degrees of freedom. The number of independent constraints is given by the
dimension of the coset space $\frac{SL(2N,\mathbb{C})\otimes SL(2N,\mathbb{C}%
)}{SL(2,\mathbb{C})}.$ We shall assume that we can use the constraints to
eliminate all degrees of freedom in $H$ and $H^{^{\prime}},$ by using the
gauge freedom to set $h_{ab}$ or $h_{ab}^{^{\prime}}$ to zero. To see this we
write the infinitesimal gauge transformations for $H$%
\[
\delta H=\omega H+H\overline{\omega}%
\]
and then evaluate it in component form, to obtain%
\begin{align*}
\delta h  &  =\omega_{l}h+h\overline{\omega}_{r}+\frac{i}{4}l^{ab}%
h_{+ab}-\frac{i}{4}h_{+ab}r^{ab},\\
\delta h_{+ab}  &  =\omega_{l}h_{+ab}+h_{+ab}\overline{\omega}_{r}+il_{+}%
^{ab}h-ihr_{+ab}\\
&  -\frac{1}{2}\left(  h_{+ac}r_{+b}^{c}-h_{+bc}r_{+a}^{c}\right)  +\frac
{1}{2}\left(  l_{+ac}h_{+b}^{c}-l_{+bc}h_{+a}^{c}\right)  ,
\end{align*}
where $l_{\pm ab}=\frac{1}{2}\left(  l_{ab}\pm i\widetilde{l}_{ab}\right)  $
and $r_{\pm ab}=\frac{1}{2}\left(  r_{ab}\pm i\widetilde{r}_{ab}\right)  $. We
can also write a similar relation for $h^{^{\prime}\text{ }}$and
$h_{+ab}^{^{\prime}}$. By fixing the gauge degrees of freedom it is possible
to set
\[
\left\langle H\right\rangle =h_{i}\lambda^{i},\quad\left\langle H^{^{\prime}%
}\right\rangle =h_{i}^{^{\prime}}\lambda^{i},
\]
where $h_{i}$ and $h_{i}^{^{\prime}}$ are constants and $\lambda^{i}$ are the
Gell-Mann $U(N)$ matrices. Obviously in this unitary gauge the action
simplifies, and the mixing terms for the matrix-valued vierbeins give masses
to all the gravitons, save one singlet, which is protected to stay massless by
diffeomorphism invariance. In the special case where $\left\langle
H\right\rangle $ and $\left\langle H^{^{\prime}}\right\rangle $ are diagonal,
the symmetry is broken down to $\left(  SL(2,\mathbb{C})\times U(1)\right)
^{N}$. Physically this is not desirable because it will imply the existence of
$N$ massless gravitons \cite{PDG}. To have a realistic theory we have to
assume that the symmetry is broken down to a $SL(2,\mathbb{C})$ so that only
one massless graviton remains. In what follows we shall study the spectrum and
show that all gravitons acquire Fierz-Pauli mass terms \cite{Fierz} through
the Higgs mechanism.

\section{The Spectrum}

To analyze the spectrum, we first derive the component form of the action. The
full action, when expressed in terms of components, is complicated and is
given in the appendix. In what follows we shall determine the spectrum by
studying perturbatively the quadratic part of the action.

To proceed, it is important to solve the torsion constraints and determine the
independent degrees of freedom. Because of the matrix nature of the equations,
these could only be solved perturbatively. To simplify matters we use the
basis of Gell-Mann matrices $\lambda^{i}$ for $U(N)$ satisfying the
commutation and anticommutation relations%
\begin{align*}
\left[  \lambda^{i},\lambda^{j}\right]   &  =2if^{ijk}\lambda^{k},\\
\left\{  \lambda^{i},\lambda^{j}\right\}   &  =2d^{ijk}\lambda^{k},
\end{align*}
and normalized by the condition $Tr\left(  \lambda^{i}\lambda^{j}\right)
=2\delta^{ij}.$ It is also important to distinguish the diagonal $U(1)$ matrix
$\lambda^{0}=\sqrt{\frac{2}{N}}1_{N}$ from the $SU(N)$ matrices $\lambda^{I},$
$I=1,\cdots,N^{2}-1$ where $i=0,I.$ Using the relations
\begin{align*}
f^{0IJ}  &  =0,\qquad d^{000}=\sqrt{\frac{2}{N}},\\
d^{0IJ}  &  =d^{I0J}=d^{IJ0}=\sqrt{\frac{2}{N}}\delta^{IJ},
\end{align*}
we can decompose the torsion equations for $E_{\mu}^{ai}$ into the form:%
\[
\partial_{\mu}E_{\nu}^{ai}+2d^{ijk}a_{\mu}^{1j}E_{\nu}^{ak}+2f^{ijk}a_{\mu
}^{2j}E_{\nu}^{ak}+d^{ijk}B_{\mu}^{abj}E_{\nu b}^{k}+f^{ijk}\widetilde{B}%
_{\mu}^{abj}E_{\nu b}^{k}-\mu\leftrightarrow\nu=0.
\]
A similar equation holds for $E_{\mu}^{^{\prime}ai}$ where the only difference
is an opposite sign for the term with $\widetilde{B}_{\mu}^{abj}$. The above
condition can be decomposed into two sets
\[
\sqrt{\frac{N}{2}}\partial_{\mu}E_{\nu}^{a0}+2a_{\mu}^{1I}E_{\nu}^{aI}+B_{\mu
}^{ab0}E_{\nu b}^{0}+B_{\mu}^{abI}E_{\nu b}^{I}-\mu\leftrightarrow\nu=0,
\]%
\[
\hspace{-2in}\partial_{\mu}E_{\nu}^{aI}+2\sqrt{\frac{2}{N}}a_{\mu}^{1I}E_{\nu
}^{a0}+2d^{IJK}a_{\mu}^{1J}E_{\nu}^{aK}+2f^{IJK}a_{\mu}^{2J}E_{\nu}%
^{aK}\allowbreak
\]%
\[
+\sqrt{\frac{2}{N}}B_{\mu}^{abI}E_{\nu b}^{0}+d^{IJK}B_{\mu}^{abJ}E_{\nu
b}^{K}+f^{IJK}\widetilde{B}_{\mu}^{abJ}E_{\nu b}^{K}-\mu\leftrightarrow\nu=0.
\]
\linebreak To solve these equations perturbatively we consider fluctuations
around a flat background. We therefore write%
\begin{align*}
E_{\mu}^{a0}  &  =\sqrt{\frac{N}{2}}\delta_{\mu}^{a}+E_{\mu}^{a0(1)}+E_{\mu
}^{a0\left(  2\right)  }+\cdots,\\
E_{\mu}^{aI}  &  =E_{\mu}^{aI\left(  1\right)  }+E_{\mu}^{aI\left(  2\right)
}+\cdots,\\
B_{\mu}^{ab0}  &  =B_{\mu}^{ab0\left(  0\right)  }+B_{\mu}^{ab0\left(
1\right)  }+B_{\mu}^{ab0\left(  2\right)  }+\cdots,\\
B_{\mu}^{abI}  &  =B_{\mu}^{abI\left(  1\right)  }+B_{\mu}^{abI\left(
2\right)  }+\cdots,\\
a_{\mu}^{I}  &  =a_{\mu}^{I\left(  1\right)  }+a_{\mu}^{I\left(  2\right)
}+\cdots.
\end{align*}
These expansions, when substituted in the above torsion constraints, could be
solved perturbatively for $B_{\mu}^{ab0}$ and $B_{\mu}^{abI}$ in terms of
$E_{\mu}^{a0}$, $E_{\mu}^{aI}$ and $a_{\mu}^{I}.$ It is possible to absorb the
corrections $E_{\mu}^{a0(1)},$ $E_{\mu}^{a0\left(  2\right)  },\cdots,$ by
using an arbitrary curved background $e_{\mu}^{a}$. Similarly we can absorb
the corrections to $a_{\mu}^{I}$ by redefining it. We therefore write%
\begin{align*}
E_{\mu}^{a0}  &  =\sqrt{\frac{N}{2}}e_{\mu}^{a},\\
a_{\mu}^{I}  &  =a_{\mu}^{I1}+ia_{\mu}^{I2},
\end{align*}
and consider these terms to be of order zero. We then have, to zeroth order%
\[
\partial_{\mu}e_{\nu}^{a}+\sqrt{\frac{2}{N}}B_{\mu}^{ab0\left(  0\right)
}e_{\nu b}-\mu\leftrightarrow\nu=0,
\]
which could be easily solved to give
\[
B_{\mu}^{ab0\left(  0\right)  }=\sqrt{\frac{N}{2}}\omega_{\mu}^{ab}\left(
e\right)  ,
\]
where $\omega_{\mu}^{ab}\left(  e\right)  $ is the usual spin-connection of
the vierbein $e_{\mu}^{a}$ .

To first order we then have
\begin{align*}
B_{\mu}^{ab0\left(  1\right)  }e_{\nu b}-\mu &  \leftrightarrow\nu=0,\\
\partial_{\mu}E_{\nu}^{aI\left(  1\right)  }+2a_{\mu}^{1I}e_{\nu}^{a}+B_{\mu
}^{abI\left(  1\right)  }e_{\nu b}+\omega_{\mu}^{ab}\left(  e\right)  E_{\nu
b}^{I\left(  1\right)  }-\mu &  \leftrightarrow\nu=0,
\end{align*}
which give the solutions
\begin{align*}
B_{\mu}^{ab0\left(  1\right)  }  &  =0,\\
B_{\mu}^{abI\left(  1\right)  }  &  =\frac{1}{2}\left(  \Omega_{\mu\nu\rho
}^{I\left(  1\right)  }-\Omega_{\nu\rho\mu}^{I\left(  1\right)  }+\Omega
_{\rho\mu\nu}^{I\left(  1\right)  }\right)  e^{a\nu}e^{b\rho},
\end{align*}
where
\[
\Omega_{\mu\nu\rho}^{I\left(  1\right)  }=D_{\mu}^{\left(  g\right)  }E_{\nu
}^{aI\left(  1\right)  }e_{a\rho}+4a_{\mu}^{1I}g_{\nu\rho}-\mu\leftrightarrow
\nu.
\]
The covariant derivative $D_{\mu}$ is taken with respect to the background
metric $g_{\mu\nu}=e_{\mu}^{a}e_{\nu a}$. This process can be continued to
second order to give the constraints%
\[
2a_{\mu}^{1I}E_{\nu}^{aI\left(  1\right)  }+B_{\mu}^{ab0\left(  2\right)
}e_{\nu b}+B_{\mu}^{abI\left(  1\right)  }E_{\nu b}^{I\left(  1\right)  }%
-\mu\leftrightarrow\nu=0,
\]

\[
\partial_{\mu}E_{\nu}^{aI\left(  2\right)  }+2d^{IJK}a_{\mu}^{1J}E_{\nu
}^{aK\left(  1\right)  }+2f^{IJK}a_{\mu}^{2J}E_{\nu}^{aK\left(  1\right)
}+B_{\mu}^{abI\left(  2\right)  }e_{\nu b}%
\]%
\[
+\omega_{\mu}^{ab}\left(  e\right)  E_{\nu b}^{I\left(  2\right)  }%
+d^{IJK}B_{\mu}^{abJ\left(  1\right)  }E_{\nu b}^{K\left(  1\right)  }%
+f^{IJK}\widetilde{B}_{\mu}^{abJ\left(  1\right)  }E_{\nu b}^{K\left(
1\right)  }-\mu\leftrightarrow\nu=0.
\]
\linebreak The solution of these equations are%
\begin{align*}
B_{\mu}^{ab0\left(  2\right)  }  &  =\frac{1}{2}\left(  \Omega_{\mu\nu\rho
}^{0(2)}-\Omega_{\nu\rho\mu}^{0(2)}+\Omega_{\rho\mu\nu}^{0(2)}\right)
e^{a\nu}e^{b\rho},\\
B_{\mu}^{abI\left(  2\right)  }  &  =\frac{1}{2}\left(  \Omega_{\mu\nu\rho
}^{I\left(  2\right)  }-\Omega_{\nu\rho\mu}^{I\left(  2\right)  }+\Omega
_{\rho\mu\nu}^{I\left(  2\right)  }\right)  e^{a\nu}e^{b\rho},
\end{align*}
where
\begin{align*}
\Omega_{\mu\nu\rho}^{0(2)}  &  =\left(  2a_{\mu}^{1I}E_{\nu}^{aI\left(
1\right)  }+B_{\mu}^{abI\left(  1\right)  }E_{\nu b}^{I\left(  1\right)
}\right)  e_{\rho a}-\mu\leftrightarrow\nu,\\
\Omega_{\mu\nu\rho}^{I\left(  2\right)  }  &  =\left(  D_{\mu}^{\left(
g\right)  }E_{\nu}^{aI\left(  2\right)  }+2a_{\mu}^{1I}e_{\nu}^{a}%
+2d^{IJK}a_{\mu}^{1J}E_{\nu}^{aK\left(  1\right)  }+2f^{IJK}a_{\mu}^{2J}%
E_{\nu}^{aK\left(  1\right)  }\quad\right. \\
&  \hspace{1in}\left.  +d^{IJK}B_{\mu}^{abJ\left(  1\right)  }E_{\nu
b}^{K\left(  1\right)  }+f^{IJK}\widetilde{B}_{\mu}^{abJ\left(  1\right)
}E_{\nu b}^{K\left(  1\right)  }-\mu\leftrightarrow\nu\right)  .
\end{align*}
Next, we solve the torsion constraints on $E_{\mu}^{^{\prime}ai}$, which give
$B_{\mu}^{abi}$ in terms of $E_{\mu}^{^{\prime}ai}$ and $a_{\mu}^{i}$. By
expanding around the background
\begin{align*}
E_{\mu}^{^{\prime}a0}  &  =\sqrt{\frac{N}{2}}e_{\mu}^{^{\prime}a},\\
\qquad E_{\mu}^{^{\prime}aI}  &  =E_{\mu}^{^{\prime}aI\left(  1\right)
}+E_{\mu}^{^{\prime}aI\left(  2\right)  }+\cdots,
\end{align*}
and equating perturbatively the two expressions obtained for $B_{\mu}^{abi}$
we deduce that%
\begin{align*}
E_{\mu}^{^{\prime}a0}  &  =\sqrt{\frac{N}{2}}e_{\mu}^{a},\qquad\\
E_{\mu}^{^{\prime}aI\left(  1\right)  }  &  =E_{\mu}^{aI\left(  1\right)  }.
\end{align*}
However, the relation between $E_{\mu}^{^{\prime}aI\left(  2\right)  }$ and
$E_{\mu}^{aI\left(  2\right)  }$ is complicated. These are $24\left(
N^{2}-1\right)  $ constraints on the $40\left(  N^{2}-1\right)  $ fields
$E_{\mu}^{^{\prime}aI\left(  2\right)  }$, $E_{\mu}^{aI\left(  2\right)  }$,
$a_{\mu}^{1I}$ and $a_{\mu}^{2I}.$ This leaves $16\left(  N^{2}-1\right)
+16=16N^{2}$ unconstrained variables corresponding to the components of a
matrix-valued vierbein.

A similar analysis can be performed on the fields $F_{\mu}^{ai}$ and $F_{\mu
}^{^{\prime}ai}$ to show that the unconstrained variables correspond to a
second matrix valued vierbein. To lowest orders, the expansions are%
\begin{align*}
F_{\mu}^{a0}  &  =\sqrt{\frac{N}{2}}f_{\mu}^{a},\\
F_{\mu}^{aI}  &  =F_{\mu}^{aI\left(  1\right)  }+F_{\mu}^{aI\left(  2\right)
}+\cdots,\\
F_{\mu}^{^{\prime}aI}  &  =F_{\mu}^{aI\left(  1\right)  }+F_{\mu}^{^{\prime
}aI\left(  2\right)  }+\cdots,\\
F_{\mu}^{^{\prime}a0}  &  =f_{\mu}^{a}+F_{\mu}^{a0(1)}+F_{\mu}^{a0\left(
2\right)  }+\cdots,\\
b_{\mu}^{I}  &  =b_{\mu}^{I1}+ib_{\mu}^{I2},
\end{align*}
where again the resulting relation between $F_{\mu}^{^{\prime}aI\left(
2\right)  }$ and $F_{\mu}^{aI\left(  2\right)  }$ is complicated.

To determine the dynamical degrees of freedom, we expand the action to terms
of order 2. This will give kinetic and mass terms. All of this could be done
in a covariant way by expanding around a curved background determined by the
two fields $e_{\mu}^{a}$ and $f_{\mu}^{a}.$ For simplicity, we group the terms
according to their perturbative order in terms of fluctuations. To zeroth
order we have%
\begin{align*}%
{\displaystyle\int\limits_{M}}
d^{4}  &  x\epsilon^{\mu\nu\kappa\lambda}\left(  \epsilon_{abcd}\left(
\left(  \lambda-\eta\right)  e_{\mu}^{a}e_{\nu}^{b}e_{\kappa}^{c}e_{\lambda
}^{d}+\left(  \lambda+\eta\right)  f_{\mu}^{a}f_{\nu}^{b}f_{\kappa}%
^{c}f_{\lambda}^{d}\right.  \right. \\
&  \qquad+C_{1}h_{i}h_{i}e_{\mu}^{a}e_{\nu}^{b}e_{\kappa}^{c}f_{\lambda}%
^{d}+C_{2}h_{i}h_{i}e_{\mu}^{a}f_{\nu}^{b}f_{\kappa}^{c}f_{\lambda}^{d}\\
&  \qquad\qquad\qquad\left.  -\frac{1}{2}\left(  \alpha_{1}-\beta_{2}\right)
e_{\mu}^{a}e_{\nu}^{b}R_{\kappa\lambda}^{cd}\left(  e\right)  +\frac{1}%
{2}\left(  \beta_{1}-\alpha_{2}\right)  f_{\mu}^{a}f_{\nu}^{b}R_{\kappa
\lambda}^{cd}\left(  f\right)  \right) \\
&  \qquad\qquad\qquad\left.  +\left(  \beta_{1}-\alpha_{2}\right)  e_{\mu}%
^{a}e_{\nu}^{b}R_{\kappa\lambda ab}\left(  e\right)  -\left(  \beta_{1}%
+\alpha_{2}\right)  f_{\mu}^{a}f_{\nu}^{b}R_{\kappa\lambda ab}\left(
f\right)  \right)  ,
\end{align*}
where
\begin{align*}
C_{1}  &  =\left(  \rho-\chi\right)  +\left(  \tau-\xi\right)  l^{2},\\
C_{2}  &  =-\left(  \rho+\chi\right)  -\left(  \tau+\xi\right)  l^{2},
\end{align*}
and for simplicity, we have assumed that%
\[
h_{i}^{^{\prime}}=lh_{i}.
\]
In reality this is full Lagrangian, in the unitary gauge, for the special case
of $N=1$. It is instructive to see the more details in this simple case. First
the gauge transformations of the Higgs fields simplify to
\begin{align*}
\delta h  &  =\frac{i}{4}\left(  l^{ab}-r^{ab}\right)  h_{+ab},\\
\delta h_{+ab}  &  =i\left(  l_{+ab}-r_{+ab}\right)  h+\frac{1}{2}\left(
l_{+ac}-r_{+ac}\right)  h_{+b}^{c}-\frac{1}{2}\left(  l_{+bc}-r_{+bc}\right)
h_{+a}^{c}%
\end{align*}
This proves that one can use the $\left(  l^{ab}-r^{ab}\right)  $ gauge
freedom to set $h_{ab}$ or $h^{^{\prime}ab}$ to zero. The constraints on the
Higgs fields then give%
\begin{align*}
hh^{^{\prime}}  &  =c_{1}+ic_{2},\\
h\partial_{\mu}h^{^{\prime}}  &  =0
\end{align*}
A solution to these equations is to have $h_{1}$, $h_{2},$ $h_{1}^{^{\prime}}$
and $h_{2}^{^{\prime}}$ as constants. The $SL(2,\mathbb{C})\times
SL(2,\mathbb{C})$ Lagrangian has all the desirable properties and gives unique
non-ambiguous interactions. This is to be contrasted with the metric theory
where the possible interactions that could be written are infinite. The last
two terms in the above action are topological. This Lagrangian gives the
interaction of one massless graviton and one massive graviton with a
Fierz-Pauli mass term \cite{Fierz}. A full analysis of a similar system was
performed many years ago and applied to a theory of massive supergravity
\cite{CSS}.

We now continue our analysis of the spectrum in the general case. We expand
the fields $e_{\mu}^{a}$ and $f_{\mu}^{a}$ around a Minkowski. background
\begin{align*}
e_{\mu a}  &  =\eta_{\mu a}+\overline{e}_{\mu a},\\
f_{\mu a}  &  =\eta_{\mu a}+\overline{f}_{\mu a},
\end{align*}
By imposing conditions that the linear fluctuations in $\overline{e}_{\mu a}$
and $\overline{f}_{\mu a}$ vanish we obtain the following relations between
the the free parameters $\lambda,$ $\eta,$ $\tau,$ $\xi,$ $\rho$ and $\chi$
\begin{align*}
\lambda &  =\frac{2}{N}\left(  \chi+\xi l^{2}\right)  h_{i}h_{i},\\
\qquad\eta &  =\frac{1}{N}\left(  \rho+\tau l^{2}\right)  h_{i}h_{i}.
\end{align*}
One then checks that using the above conditions, the cosmological constant
vanishes, and the mass terms simplify to
\[
-6\lambda%
{\displaystyle\int\limits_{M}}
d^{4}x\left(  \delta_{a}^{\mu}\delta_{b}^{\nu}-\delta_{b}^{\mu}\delta_{a}%
^{\nu}\right)  \left(  \overline{e}_{\mu}^{a}-\overline{f}_{\mu a}^{a}\right)
\left(  \overline{e}_{\nu}^{b}-\overline{f}_{\nu}^{b}\right)  ,
\]
which is of the Fierz-Pauli form. This shows that the combination $\left(
\overline{e}_{\mu}^{a}+\overline{f}_{\mu a}^{a}\right)  $ stays massless while
$\left(  \overline{e}_{\mu}^{a}-\overline{f}_{\mu a}^{a}\right)  $ acquires a mass.

Next we study the dynamical degrees of freedom associated with the vierbeins
$E_{\mu}^{aI}$ and $F_{\mu}^{aI}.$ First the linear terms are given by%
\begin{align*}
&  \frac{2}{N}%
{\displaystyle\int\limits_{M}}
d^{4}x\epsilon^{\mu\nu\rho\sigma}\epsilon_{abcd}d^{jkI}h_{j}h_{k}\left(
C_{1}\left(  3\overline{E}_{\mu}^{aI}f_{\nu}^{b}e_{\kappa}^{c}e_{\lambda}%
^{d}+\overline{F}_{\mu}^{aI}e_{\nu}^{b}e_{\kappa}^{c}e_{\lambda}^{d}\right)
\right. \\
&  \hspace{1.75in}\left.  +C_{2}\left(  \overline{E}_{\mu}^{aI}f_{\nu}%
^{b}f_{\kappa}^{c}f_{\lambda}^{d}+3\overline{F}_{\mu}^{aI}e_{\nu}^{b}%
f_{\kappa}^{c}f_{\lambda}^{d}\right)  \right)  .
\end{align*}
To eliminate this term we impose the constraint on the values $h_{i}:$%
\[
d^{jkI}h_{j}h_{k}=0.
\]
This will help reduce the mass terms for the $\overline{E}_{\mu}^{aI}$ and
$\overline{F}_{\mu}^{aI}$ fields to the form%
\begin{align*}
&  \frac{12}{N}%
{\displaystyle\int\limits_{M}}
d^{4}x\delta_{ab}^{\mu\nu}\left(  \left(  \lambda-2\eta\right)  \overline
{E}_{\mu}^{aI}\overline{E}_{\nu}^{bI}+\left(  \lambda+2\eta\right)
\overline{F}_{\mu}^{aI}\overline{F}_{\nu}^{bI}\right. \\
&  \hspace{0.9in}\left.  +2\left(  \chi+\xi l^{2}\right)  h_{j}h_{k}\left(
f_{jMP}f_{kNP}-d_{jMP}d_{kNP}\right)  \overline{E}_{\mu}^{aI}\overline{F}%
_{\nu}^{bI}\right)  .
\end{align*}
The mass terms are of the Fierz-Pauli type generated by breaking the gauge
symmetry spontaneously. For generic values of $h_{i}$ the mass matrix is
non-singular, and can be made positive definite. This way all graviton fields
$E_{\mu}^{aI}$ and $F_{\mu}^{aI}$ acquire mass terms. Note that in the special
case when the fields $H$ and $H^{^{\prime}}$ are diagonal then there will be
\ a preserved $U(1)^{N}$ subgroup of $U(N)$ and it is easy to see that the $N$
combinations $E_{\mu}^{aI}+F_{\mu}^{aI}$ corresponding to the diagonal
generators remain massless while all other fields become massive. This, of
course, is undesirable and will be avoided.

Next, we study the kinetic terms. It is well known that the above Lagrangian
give the correct kinetic terms for the fields $e_{\mu}^{a}$ and $f_{\mu}^{a}$.
We have to determine whether the fields $E_{\mu}^{aI}$ and $F_{\mu}^{aI}$
obtain the correct kinetic terms as well, and the nature of the complex
$SU(N)$ gauge fields $a_{\mu}^{I}$ and $b_{\mu}^{I}.$ First we examine the
interaction
\begin{align*}
&
{\displaystyle\int\limits_{M}}
d^{4}x\epsilon^{\mu\nu\kappa\lambda}\epsilon_{abcd}tr\left(  \left\{  E_{\mu
}^{a},E_{\nu}^{b}\right\}  B_{\kappa\lambda}^{cd}\right) \\
&  =4%
{\displaystyle\int\limits_{M}}
d^{4}x\epsilon^{\mu\nu\kappa\lambda}\epsilon_{abcd}d^{ijk}E_{\mu}^{ai}E_{\nu
}^{bj}B_{\kappa\lambda}^{cdk}\\
&  =%
{\displaystyle\int\limits_{M}}
d^{4}x\epsilon^{\mu\nu\kappa\lambda}\epsilon_{abcd}\left(  2Ne_{\mu}^{a}%
e_{\nu}^{b}R_{\kappa\lambda}^{cd}(e)+8e_{\mu}^{a}E_{\nu}^{bI}B_{\kappa\lambda
}^{cdI}+4E_{\mu}^{aI}E_{\nu}^{bI}R_{\kappa\lambda}^{cd}(e)\right)  .
\end{align*}
and concentrate on the middle term. Substituting for $B_{\mu}^{abI}$ from the
solution of the torsion constraint, we obtain, to linearized order
\begin{align*}
2B_{\mu ab}^{I}  &  =\partial_{\mu}\left(  E_{ab}^{I}-E_{ba}^{I}\right)
-\partial_{a}\left(  E_{\mu b}^{I}+E_{b\mu}^{I}\right) \\
&  +\partial_{b}\left(  E_{\mu a}^{I}+E_{a\mu}^{I}\right)  +8\left(  \eta_{\mu
a}a_{b}^{1I}-\eta_{\mu b}a_{a}^{1I}\right)  ,
\end{align*}
where indices are raised and lowered with the Minkowski metric $\eta_{ab}.$ We
also decompose $E_{\mu a}^{I}$ into symmetric and antisymmetric parts%
\[
E_{\mu a}^{I}=S_{\mu a}^{I}+T_{\mu a}^{I},
\]
where $S_{\mu a}^{I}=S_{a\mu}^{I}$ is symmetric and $T_{\mu a}^{I}=-T_{a\mu
}^{I}$ is antisymmetric. We also denote $S^{I}=\eta^{\mu a}S_{\mu a}^{I}.$ The
middle term in the above Lagrangian then gives
\begin{align*}
&  4%
{\displaystyle\int\limits_{M}}
d^{4}x\left(  \partial_{\mu}S^{I}\partial^{\mu}S^{I}-2\partial_{\mu}%
S^{I}\partial_{\nu}S^{\mu\nu I}+2\partial_{\nu}S^{\mu\nu I}\partial^{\kappa
}S_{\mu\kappa}^{I}-\partial_{\mu}S_{\nu\kappa}^{I}\partial^{\mu}S^{\nu\kappa
I}\right) \\
&  +8%
{\displaystyle\int\limits_{M}}
d^{4}xa_{\mu}^{1I}\left(  \partial^{\mu}S^{I}-\partial_{\nu}S^{\mu\nu
I}\right) \\
&  +4%
{\displaystyle\int\limits_{M}}
d^{4}x\left(  \partial_{\mu}a_{\nu}^{1I}-\partial_{\nu}a_{\mu}^{1I}\right)
T^{\mu\nu I}.
\end{align*}
The kinetic terms for the symmetric tensor $S_{\mu\nu}$ are of the standard
form for a spin-2 field \cite{Deser}. It is remarkable that the antisymmetric
part of $E_{\mu a}^{I}$ does not acquire a kinetic term, but instead couples
to the field strength of $a_{\mu}^{1I}$. Therefore, to lowest order in the
perturbative expansion, it could be considered as an auxiliary field and
eliminated. There is also the coupling%
\[%
{\displaystyle\int\limits_{M}}
d^{4}x\epsilon^{\mu\nu\kappa\lambda}e_{\mu}^{a}E_{\nu}^{bI}B_{\kappa\lambda
}^{abI},
\]
whose quadratic part simplifies to
\[
4%
{\displaystyle\int\limits_{M}}
d^{4}xT_{\mu\nu}^{I}\left(  \epsilon^{\mu\nu\kappa\lambda}\left(
\partial_{\kappa}a_{\lambda}^{1I}-\partial_{\lambda}a_{\kappa}^{1I}\right)
\right)  .
\]
We notice that $T_{\mu\nu}^{I}$ also couples to the field strength of $a_{\mu
}^{2I}$ as can be seen by expanding the term%
\[%
{\displaystyle\int\limits_{M}}
d^{4}x\epsilon^{\mu\nu\kappa\lambda}tr\left(  \left\{  E_{\mu}^{a},E_{\nu
a}^{^{\prime}}\right\}  a_{\kappa\lambda}^{2}\right)  ,
\]
whose lowest order contribution is
\[
-4%
{\displaystyle\int\limits_{M}}
d^{4}xT_{\mu\nu}^{I}\left(  \epsilon^{\mu\nu\kappa\lambda}\left(
\partial_{\kappa}a_{\lambda}^{2I}-\partial_{\lambda}a_{\kappa}^{2I}\right)
\right)  .
\]
It is then clear that by eliminating the field $T_{\mu\nu}^{I}$ both $a_{\mu
}^{1I}$ and $a_{\mu}^{2I}$ would acquire the regular $SU(N)$ Yang-Mills gauge
field strengths. A similar analysis would also apply to the antisymmetric part
of $F_{\mu a}^{I}$ giving rise to the propagation of $b_{\mu}^{1I}$ and
$b_{\mu}^{2I}.$ In reality since the quadratic terms for $E_{\mu a}^{I}$ and
$F_{\mu a}^{I}$ mix, and have to be diagonalized, the kinetic energies of
$a_{\mu}^{1I}$, $a_{\mu}^{2I},$ $b_{\mu}^{1I}$ and $b_{\mu}^{2I}$ will also
mix and have to be diagonalized as well. This shows that the fields $\left(
e_{\mu}^{a}-f_{\mu}^{a}\right)  ,$ $E_{\mu a}^{I}$, $F_{\mu a}^{I},a_{\mu
}^{1I}$, $a_{\mu}^{2I},$ $b_{\mu}^{1I}$ and $b_{\mu}^{2I}$ all have correct
kinetic energies and all obtain Fierz-Pauli mass terms after the symmetry is
broken spontaneously. The combination $\left(  e_{\mu}^{a}+f_{\mu}^{a}\right)
$ remains massless and correspond to the usual graviton. It is important to
stress that the model is based on a first order Lagrangian, where the gauge
spin-connections are determined from the zero torsion conditions to depend on
the vierbeins and their derivatives as well as on the Yang-Mills fields
$a_{\mu}$ and $b_{\mu}$. The Yang-Mills fields $a_{\mu}$ and $b_{\mu}$ will
only have first order derivatives from the gauge field strengths, but because
of the torsion constraints couple to the antisymmetric parts of the vierbeins,
which then gives them kinetic energies. In the full non-linear theory, their
will be also higher order interactions for the antisymmetric parts of the
vierbeins and therefore could not be eliminated as auxiliary fields. The
Yang-Mills fields will not have the canonical form for their kinetic energies,
but because of gauge invariance, one would expect these terms to have the
correct dynamics.

\section{Conclusions}

We have shown that it is possible to construct a sensible theory of a complex
matrix-valued graviton based on the gauge theory of $SL(2N,\mathbb{C})\otimes
SL(2N,\mathbb{C})$. Matrix-valued complex vierbeins are introduced, and in
analogy with the first order formalism of Einstein gravity, the gauge fields
are restricted by imposing torsion like constraints on the vierbeins. The
symmetry is broken down spontaneously to $SL(2,\mathbb{C})$ by using two Higgs
fields. Constraints are imposed on the Higgs fields, thus breaking the
symmetry non-linearly. In the unitary gauge the Higgs fields could be set to
constants, thus generating masses to the gravitons. Remarkably, only the
symmetric parts of the $2\left(  N^{2}-1\right)  $ fields $E_{\mu a}^{I}$ ,
$F_{\mu a}^{I}$ acquire kinetic energies, while the antisymmetric parts do
not. These, instead, couple to the complex $SU(N)\times SU(N)$ gauge fields
$a_{\mu}^{I},$ $b_{\mu}^{I}$ and after being eliminated as auxiliary fields,
give them kinetic energies. In addition there is the massless graviton
$\left(  e_{\mu}^{a}+f_{\mu}^{a}\right)  $ and a massive graviton $\left(
e_{\mu}^{a}-f_{\mu}^{a}\right)  .$ Since the masses are obtained through the
Higgs mechanism, we will not worry about the consistency of the theory of
massive gravitons. In recent works, \cite{Georgi}, \cite{spontaneous}, it was
shown that one can avoid the singularity associated with the zero mass limit
of the massive spin-2 field by giving mass to the graviton through spontaneous
symmetry breaking, and to perform the physical analysis in the non-unitary
gauge where the Higgs fields propagate.

The restriction that a metric on the manifold is not used, highly restricts
the possible terms that could be written for the Lagrangian. This also has the
added advantage that the metric of space-time is obtained dynamically after
the symmetry is broken, and is found to be neutral under the $SU(N)\times
SU(N)$ symmetry. All other massive gravitons, save for the combination
$\left(  e_{\mu}^{a}-f_{\mu}^{a}\right)  $ transform under $SU(N)\times SU(N)$
and are therefore colored. It is straightforward to couple this model to
matter, making use of vierbeins, Higgs fields and covariant derivatives. Much
work is still needed to test the consistency of this theory at higher orders
in perturbation. It will also be interesting to find out whether this
construction can arise from other formulations such as noncommutative
geometry, since the basic variables here whether vierbeins, Higgs fields or
gauge fields are all matrix-valued, and therefore noncommuting.

\section{Acknowledgments}

I would like to thank Thilbaut Damour for a critical reading of the manuscript
and for making various suggestions. I\ would also like to thank Juerg
Fr\"{o}hlich and the Institute for Theoretical Physics at ETH Z\"{u}rich for
hospitality where part of this work was done. This work is supported in part
by National Science Foundation Grant No. Phys-0313416.

\section{Appendix}

We express the different parts of the action in component form. We shall only
compute traces of Dirac matrices but not of $U(N)$ matrices. These will be
calculated in the body of the paper, but only for the quadratic part of the
Lagrangian. First we evaluate the non-Higgs terms%
\[
\frac{1}{2N}%
{\displaystyle\int\limits_{M}}
Tr\left(  i\left(  \alpha+\beta\gamma_{5}\right)  LL^{^{\prime}}F+i\left(
\overline{\alpha}+\overline{\beta}\gamma_{5}\right)  L^{^{\prime}}%
L\overline{F}+\left(  i\lambda+\gamma_{5}\eta\right)  LL^{^{\prime}%
}LL^{^{\prime}}\right)  ,
\]
which give
\begin{align*}
&  \frac{1}{N}%
{\displaystyle\int\limits_{M}}
d^{4}x\epsilon^{\mu\nu\kappa\lambda}tr\left(  \alpha_{1}^{^{\prime}}i\left[
E_{\mu}^{a},E_{\nu}^{^{\prime}a}\right]  a_{\kappa\lambda}^{1}-\alpha
_{2}^{^{\prime}}\left\{  E_{\mu}^{a},E_{\nu}^{^{\prime}a}\right\}
a_{\kappa\lambda}^{2}\right. \\
&  \qquad\qquad\qquad\quad+\beta_{1}^{^{\prime}}i\left[  F_{\mu}^{a},F_{\nu
}^{^{\prime}a}\right]  b_{\kappa\lambda}^{1}-\beta_{2}^{^{\prime}}\left\{
F_{\mu}^{a},F_{\nu}^{^{\prime}a}\right\}  b_{\kappa\lambda}^{2}\\
&  \qquad\qquad\qquad\quad+\frac{1}{2}\left(  \alpha_{1}+\beta_{2}\right)
\left(  i\left[  E_{\mu}^{a},E_{\nu}^{^{\prime}b}\right]  B_{\kappa\lambda
ab}-\frac{1}{2}\epsilon_{abcd}\left\{  E_{\mu}^{a},E_{\nu}^{^{\prime}%
b}\right\}  B_{\kappa\lambda}^{cd}\right) \\
&  \qquad\qquad\qquad\quad+\frac{1}{2}\left(  \alpha_{1}-\beta_{2}\right)
\left(  i\left[  F_{\mu}^{a},F_{\nu}^{^{\prime}b}\right]  C_{\kappa\lambda
ab}+\frac{1}{2}\epsilon_{abcd}\left\{  F_{\mu}^{a},F_{\nu}^{^{\prime}%
b}\right\}  C_{\kappa\lambda}^{cd}\right) \\
&  \qquad\qquad\qquad\quad+\frac{1}{2}\left(  -\alpha_{2}+\beta_{1}\right)
\left(  \left\{  E_{\mu}^{a},E_{\nu}^{^{\prime}b}\right\}  B_{\kappa\lambda
ab}+\frac{1}{2}\epsilon_{abcd}i\left[  E_{\mu}^{a},E_{\nu}^{^{\prime}%
b}\right]  B_{\kappa\lambda}^{cd}\right) \\
&  \qquad\qquad\qquad\quad-\frac{1}{2}\left(  \alpha_{2}+\beta_{1}\right)
\left(  \left\{  F_{\mu}^{a},F_{\nu}^{^{\prime}b}\right\}  C_{\kappa\lambda
ab}-\frac{1}{2}\epsilon_{abcd}i\left[  F_{\mu}^{a},F_{\nu}^{^{\prime}%
b}\right]  C_{\kappa\lambda}^{cd}\right) \\
&  \qquad\qquad\qquad\quad+i\left(  \lambda-\eta\right)  \left(  E_{\mu}%
^{a}E_{\nu}^{^{\prime}a}E_{\kappa}^{b}E_{\lambda}^{^{\prime}b}+E_{\mu}%
^{a}E_{\nu}^{^{\prime}b}E_{\kappa}^{b}E_{\lambda}^{^{\prime}a}-E_{\mu}%
^{a}E_{\nu}^{^{\prime}b}E_{\kappa}^{a}E_{\lambda}^{^{\prime}b}\right) \\
&  \qquad\qquad\qquad\quad+i\left(  \lambda+\eta\right)  \left(  F_{\mu}%
^{a}F_{\nu}^{^{\prime}a}F_{\kappa}^{b}F_{\lambda}^{^{\prime}b}+F_{\mu}%
^{a}F_{\nu}^{^{\prime}b}F_{\kappa}^{b}F_{\lambda}^{^{\prime}a}-F_{\mu}%
^{a}F_{\nu}^{^{\prime}b}F_{\kappa}^{a}F_{\lambda}^{^{\prime}b}\right) \\
&  \qquad\qquad\qquad\quad\left.  +\epsilon_{abcd}\left(  \left(  \lambda
-\eta\right)  E_{\mu}^{a}E_{\nu}^{^{\prime}b}E_{\kappa}^{c}E_{\lambda
}^{^{\prime}d}+\left(  \lambda+\eta\right)  F_{\mu}^{a}F_{\nu}^{^{\prime}%
b}F_{\kappa}^{c}F_{\lambda}^{^{\prime}d}\right)  \right)  ,
\end{align*}
where
\begin{align*}
\alpha_{1}^{^{\prime}}  &  =\left(  \alpha_{1}+\beta_{2}-\alpha_{2}+\beta
_{1}\right)  ,\\
\alpha_{2}^{^{\prime}}  &  =\left(  \alpha_{1}+\beta_{2}+\alpha_{2}-\beta
_{1}\right)  ,\\
\beta_{1}^{^{\prime}}  &  =\left(  \alpha_{1}-\beta_{2}-\alpha_{2}-\beta
_{1}\right)  ,\\
\beta_{2}^{^{\prime}}  &  =\left(  \alpha_{1}-\beta_{2}+\alpha_{2}+\beta
_{1}\right)  .
\end{align*}
The mixing terms
\[
\frac{1}{2N}%
{\displaystyle\int\limits_{M}}
Tr\left(  \left(  i\tau+\gamma_{5}\xi\right)  LH^{^{\prime}}LH^{^{\prime}%
}LL^{^{\prime}}+\left(  i\rho+\gamma_{5}\chi\right)  HL^{^{\prime}%
}HL^{^{\prime}}LL^{^{\prime}}\right)  ,
\]
give the contributions%
\begin{align*}
&  \frac{1}{N}%
{\displaystyle\int\limits_{M}}
d^{4}x\epsilon^{\mu\nu\kappa\lambda}\epsilon_{abcd}tr\left(  \left(  \tau
-\xi\right)  E_{\mu}^{a}H^{^{\prime}}F_{\nu}^{b}H^{^{\prime}}E_{\kappa}%
^{c}E_{\lambda}^{^{\prime}d}-\left(  \tau+\xi\right)  F_{\mu}^{a}H^{^{\prime}%
}E_{\nu}^{b}H^{^{\prime}}F_{\kappa}^{c}F_{\lambda}^{^{\prime}d}\right. \\
&  \hspace{1in}\left.  +\left(  \rho-\chi\right)  HF_{\mu}^{^{\prime}a}%
HE_{\nu}^{^{\prime}b}E_{\kappa}^{c}E_{\lambda}^{^{\prime}d}+\left(  \rho
+\chi\right)  HE_{\mu}^{^{\prime}a}HF_{\nu}^{^{\prime}b}F_{\kappa}%
^{c}F_{\lambda}^{^{\prime}d}\right) \\
&  +\frac{1}{N}%
{\displaystyle\int\limits_{M}}
d^{4}x\epsilon^{\mu\nu\kappa\lambda}\left(  \eta_{ab}\eta_{cd}-\eta_{ac}%
\eta_{bd}+\eta_{ad}\eta_{bc}\right)  tr\left(  \left(  \tau-\xi\right)
E_{\mu}^{a}H^{^{\prime}}F_{\nu}^{b}H^{^{\prime}}E_{\kappa}^{c}E_{\lambda
}^{^{\prime}d}\right. \\
&  \hspace{1in}\hspace{2in}\left.  +\left(  \tau+\xi\right)  F_{\mu}%
^{a}H^{^{\prime}}E_{\nu}^{b}H^{^{\prime}}F_{\kappa}^{c}F_{\lambda}^{^{\prime
}d}\right. \\
&  \hspace{1in}\hspace{2in}\left.  +\left(  \rho-\chi\right)  HF_{\mu
}^{^{\prime}a}HE_{\nu}^{^{\prime}b}E_{\kappa}^{c}E_{\lambda}^{^{\prime}%
d}\right. \\
&  \hspace{1in}\hspace{2in}\left.  +\left(  \rho+\chi\right)  HE_{\mu
}^{^{\prime}a}HF_{\nu}^{^{\prime}b}F_{\kappa}^{c}F_{\lambda}^{^{\prime}%
d}\right)  .
\end{align*}

\end{document}